\documentclass[11pt,a4paper]{article}
\usepackage{jheppub}
\usepackage{physics}
\usepackage{slashed}
\pdfoutput=1

\newcommand{\be}{\begin{equation}}
\newcommand{\ee}{\end{equation}}
\def\s0#1#2{\mbox{\small{$ \frac{#1}{#2} $}}}
\newcommand{\muspace}{\mkern1mu}
\newcommand{\psib}{\bar{\psi}}
\newcommand{\la}[1]{\lambda_{#1}}
\DeclareMathOperator{\sgn}{sgn}
\DeclareMathOperator{\STr}{STr}

\author{Charlie Cresswell-Hogg}
\emailAdd{c.cresswell-hogg@sussex.ac.uk}
\author{and Daniel F.~Litim}
\emailAdd{d.litim@sussex.ac.uk}
\affiliation{Department of Physics and Astronomy, University of Sussex, \\ Brighton, BN1 9QH, U.K.}

\arxivnumber{2311.16246}

\title{Scale symmetry breaking and generation of mass at quantum critical points}

\abstract{We study an asymptotically free theory of $N$ relativistic Dirac fermions and a real scalar field  coupled by Yukawa and scalar self-interactions in three dimensions using functional renormalisation. In the limit of many fermion flavours, the cubic scalar coupling becomes exactly marginal due to quantum fluctuations, leading to a line of  strongly-coupled infrared fixed points. Fermion mass can be generated  through a quantum phase transition even if chiral symmetry is absent. The line of fixed points terminates at a critical endpoint due to the loss of vacuum stability. Exactly at the endpoint, scale symmetry is broken spontaneously, leading to the generation of fermion mass. Intriguingly, the absence of chiral symmetry is a prerequisite for the spontaneous generation of fermion mass, and not a consequence thereof. We also highlight close similarities between Gross-Neveu and Gross-Neveu--Yukawa theories at and away from critical points, and establish the large-$N$ equivalence of their functional RG flows and quantum effective actions. Further implications including for conformal field theories are indicated.}

\begin{document}

\maketitle
\flushbottom

\section{Introduction}

Fixed points of the renormalisation group (RG) play an important role in quantum and statistical field theory. Ultraviolet (UV) fixed points are key for a fundamental definition of quantum field theory, while infrared (IR) fixed points relate to continuous quantum phase transitions or chiral symmetry breaking. Moreover, scale symmetry at quantum critical points often entails full conformal invariance \cite{Polchinski:1987dy}, and enforces that quantum critical theories are massless.

Interesting new effects occur if quantum scale symmetry at  critical points is broken spontaneously.
In particular, a  dilaton arises alongside mass scales  which are not determined by the fundamental parameters of the theory.
In three dimensions, the classic example are    scalar $(\phi^2)^3_{\rm 3d}$ theories at large $N$ \cite{Bardeen:1983rv,David:1984we,Litim:2017cnl,Litim:2018pxe}, where scale symmetry is broken spontaneously at the endpoint of a  line of interacting ultraviolet (UV) fixed points owing to a non-analyticity of the quantum critical potential. Subsequently, scale symmetry breaking has been observed in more complex scalar  \cite{Rabinovici:1987tf},  supersymmetric  \cite{Eyal:1986xu,Litim:2011bf,Heilmann:2012yf} and  Chern-Simons  theories coupled to matter \cite{Bardeen:2014paa,Moshe:2014bja}. More recently, the phenomenon  has   been observed at interacting UV fixed points in perturbatively non-renormalisable fermionic  $(\bar\psi\psi)^3_{\rm 3d}$ theories \cite{Cresswell-Hogg:2022lgg,Cresswell-Hogg:2022lez}. An unexpected feature  is that even though the  dynamical breaking of chiral symmetry switches on a fermion mass and chirally odd interactions, the presence of chirally odd  interactions alone does not  entail  fermion mass \cite{Cresswell-Hogg:2022lgg,Cresswell-Hogg:2022lez}. Finally, in four dimensions, it has been speculated that spontaneous scale symmetry breaking may explain the Higgs as a “light dilaton” in extensions of the Standard Model \cite{Goldberger:2008zz,Bellazzini:2012vz,Csaki:2015hcd}. This type of mechanism  may   equally arise in UV conformal theories \cite{Litim:2014uca,Bond:2017tbw,Bond:2019npq,Bond:2021tgu,Litim:2023tym} which serve as templates for model building \cite{Bond:2017wut,Hiller:2020fbu,Bause:2021prv,Hiller:2022rla}.

In this work, we investigate the phenomenon of spontaneous scale symmetry breaking and the generation of fermion mass in Gross-Neveu--Yukawa theories (GNY for short). They are asymptotically free and perturbatively renormalisable, and may develop  critical points in the IR. In condensed matter physics, GNY models often arise as effective  theories of massless Dirac fermion excitations in honeycomb lattice models \cite{Herbut:2006cs,Herbut:2009vu,Vafek:2013mpa,Janssen:2014gea,Classen:2015ssa} or topological insulators \cite{Raghu:2007ger,Vafek:2013mpa}, and are expected to belong to the same universality class as graphene  \cite{Herbut:2006cs,Hands:2008id,Herbut:2009vu,Janssen:2014gea,Classen:2015ssa,Zerf:2017zqi,Gracey:2018qba,Gracey:2018qba,Ihrig:2018hho}. Further, in the spirit of bosonisation, GNY theories are closely related to Gross-Neveu theories (GN for short) \cite{Wilson:1972cf,Eguchi:1976iz,Hasenfratz:1991it,Hands:1991py,ZinnJustin:1991yn,Hands:1992be,Karkkainen:1993ef,Weinberg:1997rv,Rosa:2000ju,Braun:2010tt} and  offer a testing ground for   spontaneous scale symmetry breaking or the naturalness of fermion mass \cite{Cresswell-Hogg:2022lgg,Cresswell-Hogg:2022lez}.

With these considerations in mind, we identify the phase structure of GNY theories including all critical points,  clarify how fermion mass is generated dynamically in the absence of chiral symmetry, and  whether spontaneous scale symmetry breaking may occur at IR critical points. The main new addition of our study are chirally odd cubic scalar self-interactions, mimicking the role played by  six-fermion interactions in GN theories. To achieve our goals, we employ  functional renormalisation in the local potential approximation, which becomes exact in the limit of many fermion flavours. We  also clarify in concrete terms the  large-$N$ duality between GN and GNY theories, both along their  functional RG flows, and on the level of  quantum effective actions.

The paper is organised as follows. After introducing GNY models and some of their key specifics (Sec.~\ref{GNY}), we provide their functional RG equations to be used throughout (Sec.~\ref{RG}).  We then identify interacting fixed points (Sec.~\ref{FP}),  scaling dimensions (Sec.~\ref{SD}), and the phase diagram including UV-IR connecting trajectories (Sec.~\ref{sec:PD}). We also find the global fixed point potentials  (Sec.~\ref{sec:potentials}) to establish the spontaneous breaking of scale symmetry and the generation of fermion mass (Sec.~\ref{SBSS}). We continue with an analysis of the large-$N$ relationship between GNY and GN theories (Sec.~\ref{sec:duality}), and close with a  discussion and conclusions (Sec.~\ref{DC}).

\section{Gross-Neveu--Yukawa theories}
\label{GNY}

We are interested in 3d euclidean quantum field theories of the GNY type, featuring $N$ flavours of four-component Dirac fermions $\psi_a$ and a single real scalar field $\phi$, with classical actions of the form
\be\label{eq:SGNY}
S_{\rm GNY} = \! \int_x \left\{ \psib_a \slashed{\partial} \psi_a \! + \tfrac12 ( \partial \phi )^2 + H  \phi  \mkern2mu \psib_a \psi_a \! + U ( \phi ) \right\} .
\ee
Interactions in these theories are parametrised by the Yukawa coupling $H$ and the scalar potential $U$. In addition to possessing a global $U(N)$ flavour symmetry,\footnote{With four-component spinors, the flavour symmetry is actually $U(N) \times U(N)$ \cite{Gehring:2015vja}.} the theories \eqref{eq:SGNY} with four-component fermions may also be invariant under the discrete ``chiral'' symmetry
\begin{align}
\nonumber 
\psi_a &\mapsto \ \ \gamma^5 \psi_a\,, \\
\label{eq:discreteSym}
 \psib_a&\mapsto -\psib_a \gamma^5\,, \\
 \nonumber
  \phi\ &\mapsto -\phi\,,
\end{align}
inherited from the four-component representation of the Clifford algebra, provided that the interaction potential $U(\phi)$ is an even function. 

Incidentally, a similar role is played by parity transformations, $x = ( x^0, x^1, x^2 ) \mapsto x' = ( x^0, -x^1, x^2 )$, with\footnote{Here we give the explicit form of the transformation in euclidean signature, where $\gamma^1$ is hermitian and squares to the identity. For lorentzian signature, the transformation rule for $\psib$ has the opposite sign.}
\begin{align}\nonumber
\psi_a ( x ) & \mapsto \ \ \gamma^1 \psi_a ( x' )\,, \\
 \label{eq:parity} \psib_a ( x ) &\mapsto -\psib_a ( x' ) \gamma^1\,,\\ \nonumber
\phi ( x )\ &\mapsto -\phi ( x' )\,,
\end{align}
and $\phi$ a pseudoscalar. This symmetry also holds in the case of two-component fermions, where the (euclidean) gamma matrices can be taken to be the Pauli matrices \cite{ZinnJustin:1991yn,Zinn-Justin:2002ecy}. Imposing invariance under either chiral \eqref{eq:discreteSym} or parity \eqref{eq:parity} transformations forbids any odd powers of $\phi$ in the potential as well as odd powers of $\psib\psi$, such as a fermion mass term. In a slight abuse of language, we will use parity and chiral symmetry interchangeably.

The GNY theory \eqref{eq:SGNY} can  also be viewed as a bosonised version of the GN  theory with relevant four-fermion (4F) interactions $\sim \la{4\rm F}(\psib_a \psi_a)^2$, whereby the scalar field $H\phi$ plays the role of $\psib_a \psi_a$ \cite{Wilson:1972cf,Eguchi:1976iz,Hasenfratz:1991it,Hands:1991py,ZinnJustin:1991yn,Hands:1992be,Karkkainen:1993ef,Weinberg:1997rv,Rosa:2000ju,Braun:2010tt}. We postpone a  detailed analysis of their interrelations at large-$N$ until  Sec.~\ref{sec:duality}.

\section{Renormalisation group}
\label{RG}

In this work, we are  interested in theories \eqref{eq:SGNY}  where chiral or parity symmetry is absent due to interactions. It is important to clarify how this impacts on the critical points, the phase structure of the theory, and the generation of fermion mass, which is no longer protected by symmetry. To that end, we investigate the theories \eqref{eq:SGNY} using functional renormalisation \cite{Polchinski:1983gv,Wetterich:1992yh,Ellwanger:1993mw,Morris:1993qb}. 

Briefly, the method  proceeds by adding a Wilsonian cutoff term to the path integral representation of the theory, bilinear in the fields, which acts to integrate out successive momentum modes of the fields from the UV to the IR. By a Legendre transform, this defines a coarse-grained effective action $\Gamma_k$, dependent on the RG scale $k$. It interpolates between a classical action $S$ at some UV scale $k = \Lambda$ and the full quantum effective action $\Gamma$, the generating functional of one-particle-irreducible correlation functions, obtained in the IR limit where all fluctuations are integrated out $(k \to 0)$. The scale dependence of $\Gamma_k$  is governed by an exact functional identity \cite{Wetterich:1992yh}
\be\label{eq:wetterich}
\partial_t \Gamma_k = \tfrac12 \STr \left\{ \big[ \Gamma^{(2)}_k + R_k \big]^{-1} \cdot \partial_t R_k \right\}
\ee
which derives  from the regulated partition function, with $t = \ln ( k / \Lambda )$. The right hand side of this equation features a functional trace in position or momentum space, as well as a trace over all internal indices. The quantity $\Gamma_k^{(2)} + R_k$ stands for the exact inverse propagator of the regulated theory and includes the cutoff function $R_k ( q )$, which provides IR regularisation of the functional trace. UV regularisation is provided by the insertion of the scale derivative of the cutoff function, which vanishes rapidly for large momenta $q$. We use optimised cutoffs \cite{Litim:2000ci,Litim:2001up,Litim:2001fd,Litim:2002cf} for both the bosonic and fermionic regulators. Other choices of cutoff constitute different RG schemes which alters non-universal aspects of RG flows but  have no impact on the physics.

Our study is based upon an ansatz for the scale dependent effective action $\Gamma_k = \Gamma_k [ \phi, \psi, \psib ]$ of the form
\be\label{eq:GammaGNY}
\Gamma_k \! = \! \! \int_x \! \left\{ Z_\psi \psib_a \slashed{\partial} \psi_a \! + \! \tfrac12 Z_\phi ( \partial \phi )^2 \! + \! H_k \phi \psib_a \psi_a \! + \! U_k ( \phi ) \right\} \! ,
\ee
where $Z_{\psi,\phi}$ are scale dependent, but not field dependent, wavefunction renormalisation factors, $H_k$ is the running Yukawa coupling, $U_k$ the running scalar potential, and $\int_x$ stands for the integral over 3d euclidean space. This ansatz with symmetric $U(\phi)$ has been widely studied in the literature \cite{Rosa:2000ju,Hofling:2002hj,Gies:2010mqh,Braun:2010tt,Scherer:2012fjq,Janssen:2014gea,Borchardt:2015rxa,Vacca:2015nta,Knorr:2016sfs}, as it contains all necessary ingredients to capture the physics of the theory's interacting IR fixed point in the chirally symmetric setting, which is equivalent to the UV fixed point of the 3d GN model \cite{Hands:1991py,ZinnJustin:1991yn,Hands:1992be,Karkkainen:1993ef,Rosa:2000ju,Braun:2010tt,Jakovac:2013jua,Jakovac:2014lqa}, and yields exact results for universal scaling dimensions in the large-$N$ limit. As we will show, this equivalence also extends to the case of explicitly broken parity by including parity-odd interactions $\sim \phi^3$  in the scalar potential. Then, \eqref{eq:GammaGNY} accurately reproduces the line of fixed points found at large $N$ in the GN model \cite{Cresswell-Hogg:2022lgg}, including its endpoints where scale symmetry is spontaneously broken \cite{Cresswell-Hogg:2022lez}.

It is convenient to introduce dimensionless, renormalised fields and couplings  scaled in units of the RG scale $k$ as
\begin{eqnarray}
\label{eq:dimensionlessField}
\sigma &=& Z_\phi^{1/2} \, \phi/k^{1/2}\,,\\
\label{eq:dimensionlessPotentials}
h &=& ( Z_\phi \,Z_\psi^2)^{-1/2} \,H_k/k^{1/2}\,,\\
u ( \sigma) &=&  U_k ( \phi )/k^{3}\,.
\end{eqnarray}
We further scale the numerical factor $A = 3 \pi^2 / (2 N)$ into the scalar field and the effective potential as $\sigma \to \sqrt{A} \sigma$ and $u \to A u$. This ensures that all scalar interactions are  normalised in units of perturbative loop factors and  powers in $N$  suitable for a large-$N$ limit. 

In the infinite-$N$ limit, the contributions of the scalar field fluctuations $\sim 1/N$ are suppressed, and the flow of the potential is driven entirely by the Yukawa interactions \cite{Braun:2010tt,Vacca:2015nta}
\begin{align}
\label{eq:flowu}
\partial_t u &= -3 u + \frac12 \left( 1 + \eta_\phi \right) \sigma \partial_\sigma u - \frac{1}{1 + (h \sigma)^2}\,.
\end{align}
While the  anomalous dimension of the fermion field vanishes   in the infinite-$N$ limit, $\eta_\psi = 0$, the  scalar field anomalous dimension, defined as $\eta_\phi = -\partial_t \ln Z_\phi \rvert_{\sigma = 0}$,\footnote{We evaluate the scalar field anomalous dimension at the background value $\sigma = 0$, which is the location of the scalar potential's minimum in all cases of interest.} can be non-trivial. Here, it takes the form
\be\label{eq:etaphi}
\eta_\phi = \tfrac52 h^2\,.
\ee
In consequence, the Yukawa interaction evolves as
\begin{align}
\label{eq:flowh}
\partial_t h &= -\tfrac12 \left( 1 - \eta_\phi \right) h\,.
\end{align}
The RG flows \eqref{eq:flowu} and \eqref{eq:flowh} with \eqref{eq:etaphi} are the central equations controlling the model.

Let us briefly comment on the interplay of chiral symmetry and the generation of fermion mass $\sim M_F \,\psib_a \psi_a$. 
In the presence of chiral symmetry \eqref{eq:discreteSym}, an explicit fermion mass term is forbidden.
If chiral symmetry is absent, a bare fermion mass, or the generation of fermion mass by fluctuations, become a possibility.  
To check whether this can happen in practice, we project the functional flow \eqref{eq:wetterich} onto the fermion mass term.
Expanding about vanishing scalar field, we find
$\partial_t m_F = -m_F$\,,
where $m_F = M_F / k$  is the dimensionless fermion mass. Evidently, fermion mass is a relevant perturbation. However,
the large-$N$ flow shows that the running is purely classical and fluctuations do not contribute to leading order in $1/N$.
We may also expand the flow around  the (dimensionless) minimum  $\sigma_0$ of the  potential where  $u'( \sigma_0)=0$, and $\la{2} = \partial_\sigma^2 u ( \sigma_0)$  the scalar mass in units of $k$. 
This corresponds to the symmetric or symmetry broken phase if the dimensionful  expectation value  of the scalar field for $k\to 0$ takes a vanishing or finite value, respectively. We find
\be\label{eq:mF2}
\partial_t m_F = -m_F \left( 1 + \frac{2 h^2}{( 1+ m_F^2 )^2 \la{2}} \right)\,.
\ee
We observe fluctuation-induced contributions to the flow, even though fermion mass remains natural in that the flow vanishes identically for vanishing mass.
Most importantly, the results establish that fermion mass cannot be generated by fluctuations alone, even in the presence of chirally odd interactions.
We  conclude that  full chiral or parity symmetry is not required to keep the fermions massless. 
Rather, the significantly milder constraint $m_F=0$ is already sufficient.

\section{Fixed points}
\label{FP}

In this section, we identify  fixed points and scaling dimensions of the GNY theory in the presence of chirally odd interactions.

From the running Yukawa interactions \eqref{eq:flowh}, we note that  fixed points are either free $(h_*=0,\eta_\phi = 0)$ or interacting $(h_*\neq 0)$ with
\be\label{eq:etaphi=1}
\eta_\phi = 1\,.
\ee
The fixed point condition states that the scalar  anomalous dimension must exactly cancel the canonical mass dimension of the Yukawa coupling. This entails that the mass dimension of the Yukawa coupling is transfered to the scalar field, which now scales anomalously with mass dimension unity rather than $\s012$. Together with   \eqref{eq:etaphi}, this also determines the critical Yukawa coupling as
\be\label{eq:hFP}
h_*^2 = \tfrac25\,.
\ee
Notice that the fixed point value depends on the RG scheme and the choice for the momentum cutoff function $R_k$, whereas the existence of the fixed point and its scaling dimension is independent thereof. Also, the sign of $h$ is physically irrelevant and we take $h\geq0$ without loss of generality. 
Together with \eqref{eq:etaphi} and in terms of $h^2_\Lambda=h^2(k=\Lambda)$,   the Yukawa flow can be integrated explicitly to give
\be\label{hflow}
h^2(k)=h^2_\Lambda\,{\left[\frac{h_\Lambda^2}{h^2_*}+\frac{k}{\Lambda} \left(1-\frac{h_\Lambda^2}{h^2_*} \right)\right]^{-1}}\,.
\ee
The flow  interpolates  between the asymptotically free fixed point in the UV ($k \to\Lambda$) and the fixed point \eqref{eq:hFP} in the IR ($k \to 0$).
Provided that the initial Yukawa coupling takes values in the range
\be
0 < h_\Lambda^2 < h_*^2\,,
\ee
the high scale can be removed $(\Lambda\to \infty)$. The transition from the UV  to the IR fixed point is characterised by the RG invariant cross-over scale
\be
k_{\rm cr}=\Lambda\, h^2_\Lambda/h^2_*
\ee
with $\Lambda\s0{d}{d\Lambda} k_{\rm cr}=0$, which arises from dimensional transmutation close to the UV fixed point ($h^2_\Lambda\ll 1$). By extension, the anomalous dimension  then interpolates between $\eta_\phi = 0$ in the UV and $\eta_\phi = 1$ in the IR. 

In turn, provided the initial value of the Yukawa coupling $h^2_\Lambda$ exceeds its fixed point value $h^2_*$, the running coupling invariably
exhibits a Landau pole at the scale 
\be\label{eq:kPole}
{k_L} = \frac{\Lambda\,h_\Lambda^2}{h_\Lambda^2 - h_*^2}\,.
\ee
In this case,  the flow is not connected to the free UV fixed point, the high scale $\Lambda$ cannot be removed ($k_L\ge \Lambda$), and the theory is effective rather than fundamental.

Next, we turn our attention to the self-interactions of the scalar field. To this end, 
we expand the scalar potential in terms of $n$-scalar self-couplings $\la{n}$ at vanishing field
\be\label{eq:polynomialExpansion}
u ( \sigma) = \sum_{n=0}^\infty \tfrac{1}{n!} \,\la{n}\,\sigma^n\,,
\ee
allowing both even and odd interactions under the discrete symmetry \eqref{eq:discreteSym}. Inserting the series into the flow equation \eqref{eq:flowu} yields RG equations for  $\la{n}$. For the dimensionless scalar mass term ($\la2\equiv m_s^2$) we find
\be\label{eq:flowScalarMass}
\partial_t \la2 = \left( -2 + \eta_\phi \right) \la2 + 2 h^2,
\ee
with the anomalous dimension $\eta_\phi$ given by \eqref{eq:etaphi}. Notice that the scalar mass is not natural in that it will always be switched on by Yukawa interactions, regardless of its initial value. Integrating \eqref{eq:flowScalarMass} together with \eqref{hflow}, and  using $\la{2,\Lambda}=\la{2}(k=\Lambda)$, we find
\be\label{la2k}
\la{2}(k)=2h^2(k)\left[1+\left(\frac{\la{2,\Lambda}}{2h^2_\Lambda}-1\right)\frac{\Lambda}{k}\right]\,.
\ee
Provided the UV initial conditions obey $\la{2,\Lambda}>2h^2_\Lambda$, the second term in \eqref{la2k} dominates the IR and we conclude that the physical scalar mass $M^2_s\equiv\la{2}(k)k^{2-\eta_\phi}$  is positive and finite for $k\to 0$, giving a disordered phase with a vanishing vacuum expectation value. By the same token, $\la{2,\Lambda}<2h^2_\Lambda$ implies a massive, ordered phase with a non-vanishing vacuum expectation value. We conclude that the short-distance initial condition
\be\label{boundary}
\la{2,\Lambda}=2h^2_\Lambda
\ee 
defines the boundary between an ordered and disordered phase, which is also a necessary condition for a massless scalar field in the IR. We come back to this aspect in Sec.~\ref{sec:PD}.

Next, we turn to the chirally odd cubic scalar self-interaction $\lambda_3$, whose flow is given by
\be\label{eq:flow-lambda3}
\partial_t \la{3} = -\tfrac32(1 - \eta_\phi ) \la{3}\,.
\ee
The flow for $\la{3}$ is natural in that  the coupling cannot be switched on if set to zero at any one scale. 
Further, the flow \eqref{eq:flow-lambda3} vanishes identically as soon as $ \eta_\phi =1$, irrespective of the value for $\la{3}$. 
Stated differently, the Yukawa coupling taking an interacting fixed point, \eqref{eq:etaphi=1}, invariably turns the cubic self-interaction $\la{3}$ into an exactly marginal coupling. Thus, the canonical mass dimension $[\la{3}]=\tfrac32$  of the classically relevant interaction has been exactly marginalised quantum mechanically owing to strong fluctuations in the IR. This non-perturbative effect entails that the cubic coupling becomes  a free parameter,
\be\label{eq:l3freeParam}
\la{3}^* = \text{free parameter},
\ee
characterising a line of interacting IR fixed points.
Integrating the  flow for the cubic coupling \eqref{eq:flow-lambda3} and also using \eqref{hflow}, we find
\be\label{la3k}
\la{3}(k)=\la{3,\Lambda}\,h^3(k)/h^3_\Lambda\,,
\ee
where $\la{3,\Lambda}=\la{3}(k=\Lambda)$. From \eqref{la3k}, we observe that the exactly marginal parameter  \eqref{eq:l3freeParam} at the IR fixed point can be traced back to   initial conditions at the high scale as $\la{3}^*=\la{3,\Lambda} \,h_*^3/h_\Lambda^{3}$.

Turning to the quartic and higher order scalar self-interaction, we find
\be\label{eq:flow-lambda}
\partial_t \la{n} = \left[ -3 + \tfrac{n}{2} \left( 1 + \eta_\phi \right) \right] \la{n} - a_n h^{n}\,,
\ee
with coefficients $a_{n={\rm even}}=(-1)^{n/2}\, n!$ and $a_{n={\rm odd}}=0$. 
It follows that all chirally odd self-interactions $\la{n={\rm odd}}$ are natural and admit free fixed points $\la{n={\rm odd}}^*=0$, irrespective of what happens in the Yukawa sector, though with the exception of the cubic interaction which  may become exactly marginal provided $h_*\neq 0$. On the other hand, all chirally even interactions $\la{n={\rm even}}$  take free fixed points provided that $h_*=0$, and non-trivial ones with
\be\label{eq:FPcouplings}
\la{n}^* = \frac{a_n h_*^n}{n - 3}\,,
\ee
provided $h_*\neq 0$. Altogether, we conclude that the theory admits two types of fixed points, the Gaussian fixed point 
$(h_*=0,\la{n}^*=0)$ where all couplings vanish identically, and a line of interacting IR fixed points characterised by \eqref{eq:hFP}, \eqref{eq:l3freeParam} and \eqref{eq:FPcouplings}. Notice also that only the cubic scalar self-interaction $\la{3}^*\neq 0$ breaks the discrete symmetry \eqref{eq:discreteSym} at the interacting fixed point.

\section{Scaling dimensions}
\label{SD}

Next, we turn to the universal scaling dimensions. In the vicinity of a fixed point, the RG flow is characterised by a set of RG eigenoperators $\mathcal{O}_n \sim k^{3+\vartheta_n}$, which scale with universal exponents $\vartheta_n$. Positive exponents are associated to irrelevant operators, whose contributions shrink towards the IR and do not affect the long distance physics, while negative exponents indicate relevant operators. At a UV fixed point, the latter are in one-to-one correspondence with the fundamentally free parameters of the theory.

\begin{figure}
\centering
\includegraphics[width=.6\linewidth]{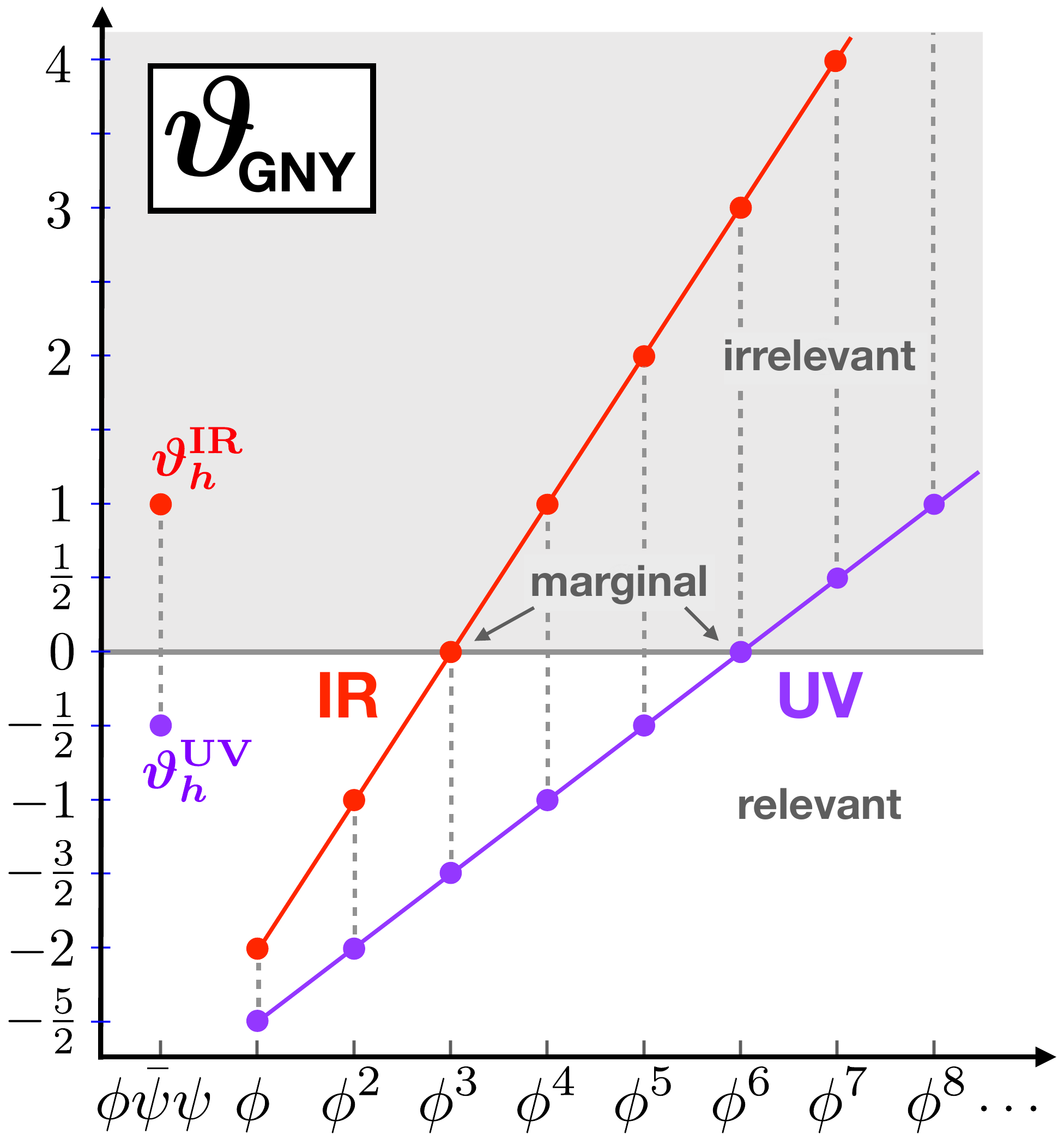}
\caption{Shown are  universal scaling exponents of the GNY theory at the UV and IR fixed points, and their association to field monomials (solid lines are added to guide the eye). Scaling exponents are shifted upwards in the IR due to quantum effects dictated by  \eqref{eq:etaphi=1}.} 
\label{fig:pThetas}
\end{figure}

At the asymptotically free UV fixed point, basic scaling operators are given by field monomials such as $\phi^n$, $(\psib\psi)^m$,  and combinations  $\phi^n (\psib\psi)^m$ thereof. Scaling exponents are mostly determined by canonical power counting. 
Relevant perturbations  
are identified as the mass terms, the Yukawa interaction $\sim h\,\phi \psib\psi$ with exponent 
\be \vartheta^{\rm UV}_h = -\tfrac 12\,,\ee 
and scalar interactions $\sim \la{n}\phi^n$ with exponents 
\be \vartheta^{\rm UV}_n = -3 + \tfrac n2\,.
\ee
Of these, all interactions with $n\le 5$ are relevant. The exponent $\vartheta_6 = 0$ relates to the classically marginal sextic interaction $\phi^6$. All higher order scalar self-interactions with $n > 6$ are irrelevant  with $\vartheta^{\rm UV}_n >0$. Our results for UV scaling exponents of scalar field monomials are illustrated in Fig.~\ref{fig:pThetas} (magenta dots).

At the interacting IR fixed points, quantum fluctuations and strong coupling effects modify the scaling behaviour.
Eigenoperators are found by linearising the flow around the fixed point. 
We find that the Yukawa interaction has become irrelevant with 
\be \label{thetaY} \vartheta^{\rm IR}_h = 1\,.\ee 
The scaling of the  scalar self-interactions is  modified by the anomalous dimension, giving 
\be \label{thetaGNY} \vartheta^{\rm IR}_n = -3 + n\,.\ee 
Results confirm that $\phi^3$ interactions have become exactly marginal, \eqref{eq:l3freeParam}, with $\la{3}$  parametrising the line of IR fixed points. Overall, only  the scalar and fermion mass terms remain relevant perturbations in the IR. 
Our results for IR scaling exponents of scalar field monomials are illustrated in Fig.~\ref{fig:pThetas} (red dots). We observe that quantum effects have induced large corrections to  scaling exponents dictated by \eqref{eq:etaphi=1}.

Switching on any of the relevant perturbations in the UV initiates a flow  towards the IR. 
The set of RG trajectories running out of the UV fixed point is then characterised by small deviations $\delta \alpha(\Lambda)=\alpha(\Lambda)-\alpha_*$ from the (free) fixed point for any of the relevant or marginal couplings $\alpha\in \{ m_F, \la{2}\equiv m_s^2,\la3,\la4,h,\la5,\la6\}$ at the high scale $\Lambda$. Below, we establish in \eqref{eq:sep} that a tuned combination of Yukawa and scalar mass perturbations  guarantee that the interacting IR fixed points are reached.

\section{Phase diagram}
\label{sec:PD}

In the UV, the theory displays an asymptotically free Gaussian fixed point where relevant perturbations  trigger RG flows towards the IR. For these, specific choices need to be made in order to reach the line of interacting fixed points in the IR. A central role is played by the Yukawa interaction, whose RG flow is given by \eqref{hflow}.

Next, we identify the UV-IR connecting separatrices. For that, the fermion mass does not play any role as it can safely be set to zero in the UV. 
However, since the scalar mass term $m_s^2 \equiv \la{2}$ is relevant at the IR fixed points, its UV initial value must be tuned precisely  to guarantee that the condition $\la{2}^* = 2 h_*^2$ is reached in the IR, see \eqref{la2k}, \eqref{boundary}. The condition for this to happen  can also be understood by considering the ratio of couplings $X=(2h^2)/\la{2}$. Using \eqref{eq:flowh} and \eqref{eq:flowScalarMass} we find
\be\label{eq:flowX}
\partial_t X = X(1-X)\,.
\ee
The flow has a free IR fixed point $X_*=0$ and an interacting UV fixed point $X_*=1$, independent of $\eta_\phi$. Therefore, the condition $\la{2} \equiv 2 h^2$ can only be achieved in the IR limit provided that it already holds true for all scales. This demonstrates that the relation
\be \label{eq:sep}
\la{2}(k) = 2\, h^2(k)
\ee 
between the scalar mass and the Yukawa coupling  identifies all  separatrices connecting the UV and IR fixed points, for all scales $k\leq \Lambda$, in accord with \eqref{la2k},  \eqref{boundary}. Accordingly, the required initial condition for relevant perturbations in the UV read $\delta \la{2}(\Lambda) = 2\, \delta h^2(\Lambda)$.

By the same token, UV initial conditions of the form $\delta \la{2}(\Lambda) = 2\, \delta h^2(\Lambda)+\epsilon(\Lambda)$ which deviate from \eqref{eq:sep}, cannot reach the line of massless IR fixed points. Instead, we find different types of massive theories in the IR, depending on the sign of $\epsilon$. For $\epsilon(\Lambda) > 0$, the IR theory contains a massive scalar, whereas the fermions remain massless. For $\epsilon(\Lambda)< 0$, the scalar potential develops a nontrivial minimum, implying that the fermions also become massive due to Yukawa interactions. This pattern holds true  regardless of whether or not chiral symmetry is explicitly broken by cubic scalar interactions.

\begin{figure}
\centering
\includegraphics[width=.6\linewidth]{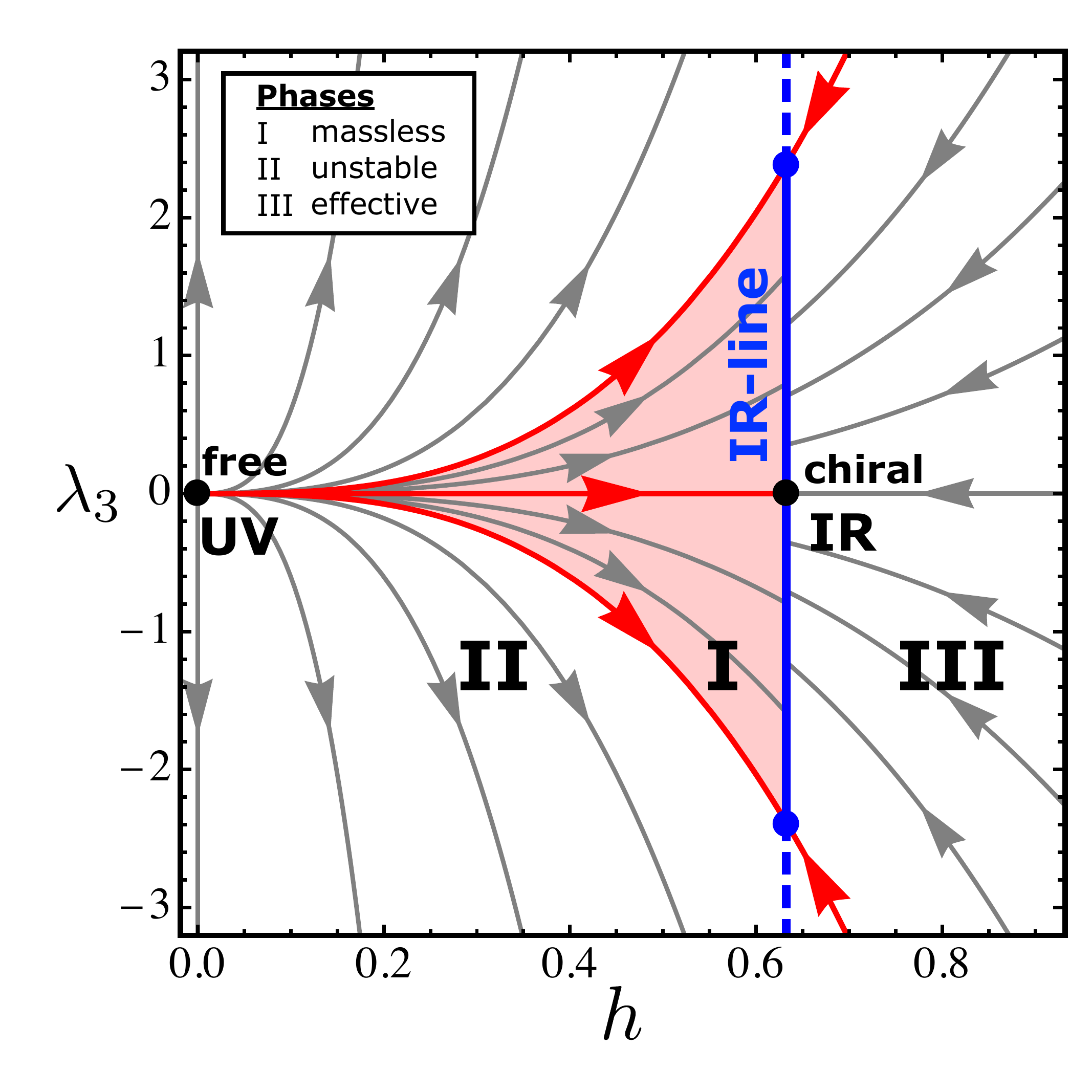}
\caption{Phase diagram of the GNY theory in terms of the Yukawa $h$ and cubic  coupling $\la 3$, and projected onto the hypersurface  \eqref{eq:sep}. We observe the free UV fixed point (black), a line of interacting IR fixed points with stable (blue), or unstable ground states (dashed blue), UV-IR connecting separatrices (red) and sample trajectories (gray). In region I (region III), we observe massless fundamental (effective) theories which (do not) originate out of an UV fixed point. In region II, theories are driven into vacuum instability; arrows point from the UV to the IR.}
\label{fig:phaseDiagram}
\end{figure}

We are now in a position to discuss the phase diagram of the theory with cubic interactions. The flow of the scalar mass does not influence this coupling, but we assume for argument's sake that $m_s^2$ has been tuned to lie on the separatrix trajectory. Figure~\ref{fig:phaseDiagram} depicts the RG flow in the $(h, \la{3})$ plane. Selected RG trajectories are shown, with arrows pointing from UV to IR. Highlighted on the left of the diagram is the UV free fixed point, which is connected via RG trajectories to the IR line of interacting fixed points, highlighted in blue on the right of the diagram. At the centre of the IR line, with $\la{3} = 0$, the chirally symmetric fixed point is marked ``chiral''. Endpoints are marked on the line to show the range of cubic coupling values in the IR for which the scalar potential is bounded from below. 

The phase diagram is divided into three regions. Region I consists of well-defined asymptotically free trajectories which flow towards interacting conformal theories with stable vacua in the IR. For these theories, all fields  remain massless at all scales, despite the fact that chiral symmetry is explicitly broken by the cubic interaction. Region II also comprises  asymptotically free trajectories, except that these are unphysical owing to an unstable vacuum in the IR limit. Region III consists of well-defined effective theories  whose running couplings are not UV complete because they exhibit UV Landau poles. Outside the $(h, \la{3})$ plane, mass can be generated either explicitly, or dynamically via  a non-trivial expectation value for the scalar field, or spontaneously, via the breaking of quantum scale symmetry.

\section{Global fixed points}
\label{sec:potentials}

In this section, we discuss the global form of the scalar potential $u_* ( \sigma )$ at the IR fixed points of the flow equation \eqref{eq:flowu} for all values of the field $\sigma$. The results we present in this section arise as a special case of those in App.~B of \cite{Vacca:2015nta} and agree with \cite{Braun:2010tt} in the chiral limit. Setting the Yukawa coupling $h$ to its IR fixed point value \eqref{eq:hFP} where $\eta_\phi = 1$, and solving the fixed point condition $\partial_t u = 0$ for the entire scalar  potential, we find
\be\label{eq:FPsolutions}
u_* ( \sigma ) = \s0{1}{6}\la{3}\, \muspace \sigma^3 + F ( h_* \sigma )\,,
\ee
with $F ( x ) = -\frac13 + x^2 + x^3 \arctan ( x )$ an even function in $x$,  and $\la{3}$ the value of the exactly marginal cubic scalar coupling at the fixed point. Expanding \eqref{eq:FPsolutions} in powers of $\sigma$, we recover all couplings $\la{n}^*$ determined previously from \eqref{eq:flow-lambda}, \eqref{eq:FPcouplings}. Given that all parity-odd couplings bar $\la{3}$ vanish at the fixed point, $\la{n={\rm odd}\neq3}=0$, it follows that  the non-perturbative function $F$ resums all parity-even polynomial couplings.  

Because $F$ is an even function, and because the Yukawa interaction is invariant under \eqref{eq:discreteSym}, the fixed point theory represented by \eqref{eq:FPsolutions} with $\la{3} = 0$ is parity symmetric. For any $\la{3}\neq 0$, however, parity symmetry is explicitly broken by the appearance of a cubic term in the scalar potential. Note that the cubic coupling has mass dimension $\s032$ at the classical level. Its exact marginality in the IR originates from the anomalous scaling of the scalar field, which effectively acquires a mass dimension of unity at the interacting fixed points.

We emphasize that the cubic interaction term in the potential may destabilise the theory. This is so because the fluctuation-induced function $F(x)$ in \eqref{eq:FPsolutions} behaves asymptotically as $\abs{x}^3$ for large arguments. As a result, the parity-odd and parity-even terms in \eqref{eq:FPsolutions} compete and the effective potential may become unbounded from below if the parameter $\la{3}$ becomes large enough. More explicitly, for large fields, the potential behaves as
\be\label{ustar}
u_* ( \sigma ) \sim \s01{6}\left[\la{3} + \la{3,\rm crit}\sgn ( \sigma ) \right]\sigma^3,
\ee
where we have introduced the critical parameter
\be\label{crit}
\la{3,\rm crit}\equiv 3\pi \,h_*^3\,.
\ee
Notice that the contribution $\sim \la{3}\sigma^3$ in \eqref{ustar} arises from the parity-odd cubic interaction whereas the contribution $\sim \la{3,\rm crit}\sgn ( \sigma ) \sigma^3$ arises from the sum of all polynomial and parity-even scalar self interactions. The effective potential found in a mean field analysis of the chirally symmetric theory has such a form \cite{Rosenstein:1988pt,Rosenstein:1988dj,Scherer:2012fjq}. From the explicit expression \eqref{ustar}, we conclude that the potential is bounded from below provided that 
\be\label{stable}
\abs{\la{3}} \leq \la{3,\rm crit}\,,
\ee
as indicated in Fig.~\ref{fig:phaseDiagram}, but elsewise unbounded for large positive ($\la{3}<-\la{3,\rm crit}$) or negative ($\la{3}>\la{3,\rm crit}$) fields, respectively. In the special case $\abs{\la{3}} = \la{3,\rm crit}$, parity-even and parity-odd contributions cancel out exactly for asymptotically large positive (negative) fields.

\begin{figure}
\centering
\includegraphics[width=.6\linewidth]{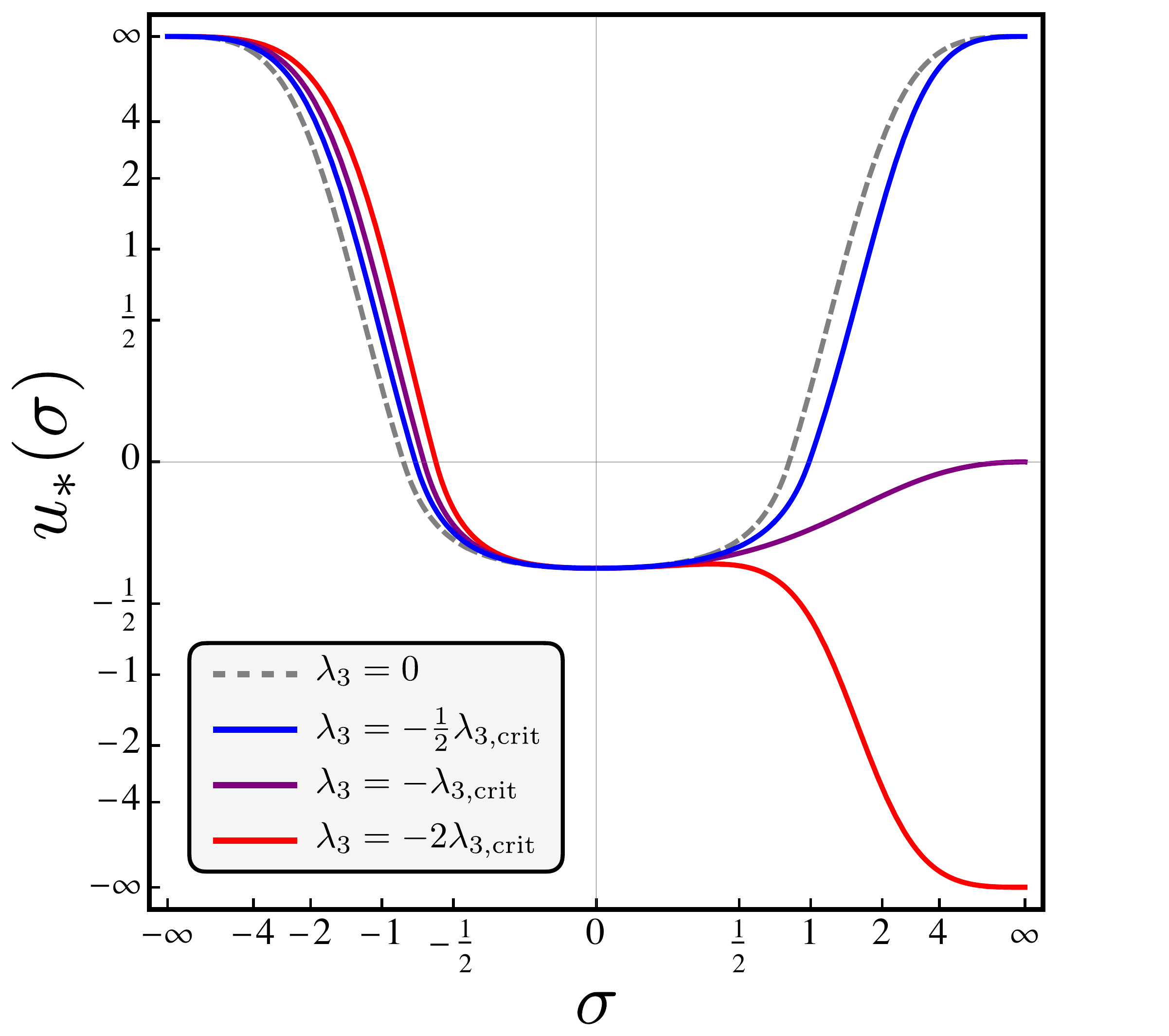}
\caption{Shown are dimensionless global potentials at IR critical points \eqref{eq:FPsolutions} for different values of $\la{3}$, covering the parity-even potential ($\la{3} = 0$, dashed grey curve), and fixed points with $\la{3} < 0$ including a stable potential (solid blue), an asymptotically vanishing  potential  (purple), and an unbounded potential (red). Examples with $\la{3} \to -\la{3}$ are obtained by reflection about the vertical axis. Axes are scaled as $X \to X / ( 1 + \abs{X} )$ for better display.}
\label{fig:FPpotentialsIR}
\end{figure}

Results are illustrated in Fig.~\ref{fig:FPpotentialsIR}, showing the fixed point potential for the parity-symmetric theory ($\la{3} = 0$, dashed grey), as well as a (bounded) parity-asymmetric potential ($\la{3} = -\frac12 \la{3, \rm crit}$, blue), an unbounded potential ($\la{3} = -2 \la{3,\rm crit}$, red), and an asymptotically flat potential ($\la{3} = -\la{3,\rm crit}$, purple). By symmetry, and given that $F$ in \eqref{eq:FPsolutions} is even, it follows that critical potentials with parameter $\la{3}\leftrightarrow -\la{3}$ relate to a  reflection in field space $\sigma \leftrightarrow -\sigma$. Lastly, we emphasize that the loss of vacuum stability beyond the bound \eqref{crit}, \eqref{stable} is a non-perturbative effect. It required the precise knowledge of the potential at asymptotically large fields, \eqref{ustar}, for which any finite order polynomial approximation of \eqref{eq:FPsolutions} would not have been sufficient.

\section{Spontaneously broken scale invariance and fermion mass}
\label{SBSS}

The critical theories  discussed in the previous sections, characterised by the value of the exactly marginal cubic coupling, are all scale invariant at the quantum level. Provided that the cubic coupling is not too large in magnitude, $\abs{\la{3}} \leq \la{3,\rm crit}$, these theories also possess stable ground states. In this section, we discuss how fixed points  at the borderline of vacuum stability ($\abs{\la{3}} = \la{3,\rm crit}$) give rise to the spontaneous breaking of quantum scale symmetry and the generation of fermion mass.

To make this  explicit, we recast  fixed point potentials \eqref{eq:FPsolutions} in terms of the dimensionful potential and fields. 
Recall that the scalar field scales anomalously at the IR fixed point, with $\eta_\phi|_* = -\partial_t \ln Z_\phi|_* =1$. The anomalous scaling  changes the mass dimension of the  scalar field by $\s012$,
corresponding to $Z_\phi|_* = {k_0}/{k}$ with $k_0$ a suitable reference scale.
Therefore, and in view of \eqref{eq:dimensionlessField}, it is convenient to introduce a new field $\varphi$ with canonical mass dimension one,  $\varphi=\sqrt{k_0} \muspace \phi \equiv  k \muspace \sigma$, and to express the dimensionful fixed point potentials as $U_k^* ( \varphi ) = k^3 u_* ( \varphi/k )$.
In the limit where all  fluctuations are integrated out ($k \to 0$) we find the full quantum effective potential as
\be\label{eq:k0potentials}
U_0^* ( \varphi ) = \s016\left[ \la{3} + \la{3,\rm crit} \sgn ( \varphi ) \right] \varphi^3\,.
\ee
We emphasize that the theory sits precisely at the IR fixed point. From the explicit result, it follows that the potential is globally unstable outside the range \eqref{stable}.

\begin{figure}
\centering
\includegraphics[width=.6\linewidth]{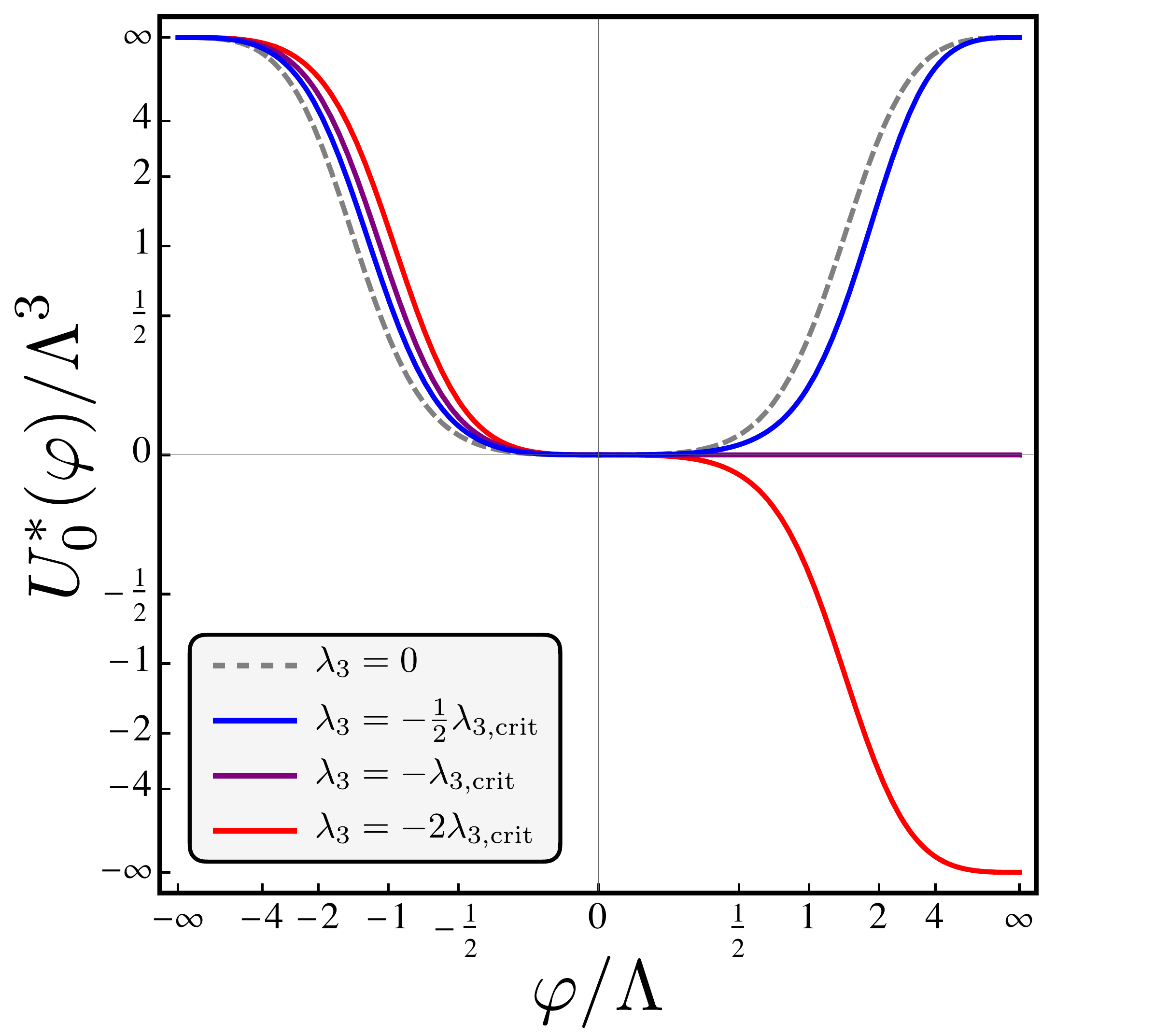}
\caption{Shown are dimensionful fixed point potentials $U_*(\varphi)$ at $k = 0$ in units of the high scale $\Lambda$ and for different  $\la{3}$, covering the parity symmetric case ($\la{3}=0$, dashed grey curve), and examples of fixed point potentials with $\la{3} < 0$ with a stable  vacuum (solid blue), flat potential and spontaneously broken scale symmetry (purple), and an unbounded potential (red). Potentials are stable for $\abs{\la{3}} \leq  \la{3,\rm crit}$; axes are scaled as in Fig.~\ref{fig:FPpotentialsIR}.}
\label{fig:k0potentialsIR}
\end{figure}

Figure~\ref{fig:k0potentialsIR} illustrates the fixed point potentials in \eqref{eq:k0potentials} for several values of the free parameter $\la{3}$ (corresponding to their dimensionless counterparts  in Fig.~\ref{fig:FPpotentialsIR}).
Most notably, spontaneous breaking of scale invariance is evidenced in the flatness of the critical potentials
with $\abs{\la{3}} = \la{3,\rm crit}$. 
Here, the minima are  degenerate, covering the entire positive (negative) field axis provided $\la{3}=-\la{3,\rm crit}$ $(\la{3}=\la{3,\rm crit})$. Using \eqref{eq:k0potentials}, the vacuum expectation value is determined from the condition
\be
\left[ \la{3} + \la{3,\rm crit} \sgn ( \varphi_0 )  \right] \varphi_0^2 = 0\,,
\ee
valid in the range \eqref{stable}. It establishes that the vacuum expectation value of the scalar field either vanishes identically as long as  $\abs{\la{3}} < \la{3,\rm crit} $, or, provided $\abs{\la{3}} = \la{3,\rm crit}$, turns into a free parameter of the theory, 
\be\label{free} 
\abs{\varphi_0} = \text{free parameter}
\ee
up to its sign which is determined by the sign of $\la{3}$. Most importantly, a non-trivial vacuum expectation value \eqref{free} spontaneously introduces a mass scale despite of the fact that the theory is tuned to a quantum critical point. The fermions thereby acquire a mass
\be\label{MF}
M_\psi = h_*\, \varphi_0
\ee 
owing to the Yukawa interactions, while the scalar remains strictly massless owing to the flatness of the critical potential, irrespective of $\varphi_0$. Thus, quantum scale symmetry is broken spontaneously, leading to the generation of fermion mass. The breaking of scale symmetry also entails a dilaton, which is expected to be massless at large $N$. 
 
As a final comment, we stress that for the above mechanism to be operative, the absence of parity symmetry is a prerequisite and not a consequence. Therefore, fermion mass is generated at a critical point, spontaneously, without the breaking of a symmetry other than scale symmetry itself.

\section{Duality with Gross-Neveu theory}
\label{sec:duality}

It is often argued that the  IR fixed point of the chirally symmetric and perturbatively renormalisable GNY theory can be viewed as the UV completion of the  perturbatively non-renormalisable GN theory with fundamental 4F interactions \cite{Wilson:1972cf,Hands:1991py,ZinnJustin:1991yn,Hands:1992be,Karkkainen:1993ef,Rosa:2000ju}. In this section, we highlight similarities between GNY \eqref{eq:SGNY} and GN theories \cite{Cresswell-Hogg:2022lgg,Cresswell-Hogg:2022lez} beyond chiral symmetry, at and away from critical points, and establish the large-$N$ equivalence of RG flows and quantum effective actions.

The action for the 3d GN theory of interacting fermions $\psi_a$ can be written as
\be\label{eq:SGN}
S_{\rm GN} = \int_x \left\{ \psib_a \slashed{\partial} \psi_a \! + V(\psib_a \psi_a) \right\} .
\ee
The function $V(z)$  (with $z=\psib_a \psi_a)$ contains  interaction monomials $\sim \la{2n\rm F}\,z^n$. The fermion anomalous dimension can be neglected in the large-$N$ limit of many fermion flavours adopted here. The theory is chirally symmetric if all $\la{2n\rm F}=0$ for $n=\rm odd$.

The theory has been studied using functional renormalisation \eqref{eq:wetterich} in \cite{Jakovac:2013jua,Cresswell-Hogg:2022lgg,Cresswell-Hogg:2022lez}. Even though the theory \eqref{eq:SGN} is non-renormalisable in perturbation theory, it is non-perturbatively renormalisable and predictive up to highest energies,  curtesy of an interacting UV fixed point \cite{Wilson:1972cf,Gawedzki:1985ed,Gawedzki:1985jn,Rosenstein:1988pt,deCalan:1991km,Hands:1991py,Gat:1991bf,Hands:1992be,Braun:2010tt,Jakovac:2014lqa}. Thereby, the classically irrelevant 4F coupling $\la{4\rm F}$ becomes relevant in the UV due to strong quantum effects. At the same time, the parity-odd $\la{6\rm F}$ becomes exactly marginal, while all other couplings $\la{n\rm F}$ with $n>6$ remain irrelevant \cite{Cresswell-Hogg:2022lgg}. Consequently, the theory displays a line of UV fixed points characterised by the exactly marginal coupling $\la{6\rm F}$. The critical line is found to be finite, \be \label{6Fcrit} \abs{\la{6\rm F}}\le \la{6\rm F, crit}\,,\ee because  a well-defined  quantum effective action ceases to exist for large sextic coupling \cite{Cresswell-Hogg:2022lgg,Cresswell-Hogg:2022lez}. At the endpoint, quantum scale symmetry is broken spontaneously, leading to the generation of a fermion mass and a massless dilaton \cite{Cresswell-Hogg:2022lez}. The critical 6F coupling reads
\be\label{eq:6Fcrit}
\la{6 \rm F, crit} = {3 \pi}/{8}\,,
\ee
if we adopt the same RG scheme and fermionic regulator function as in this work  \cite{Litim:2000ci,Litim:2001up,Litim:2002cf}.

\begin{figure}
\centering
\includegraphics[width=.6\linewidth]{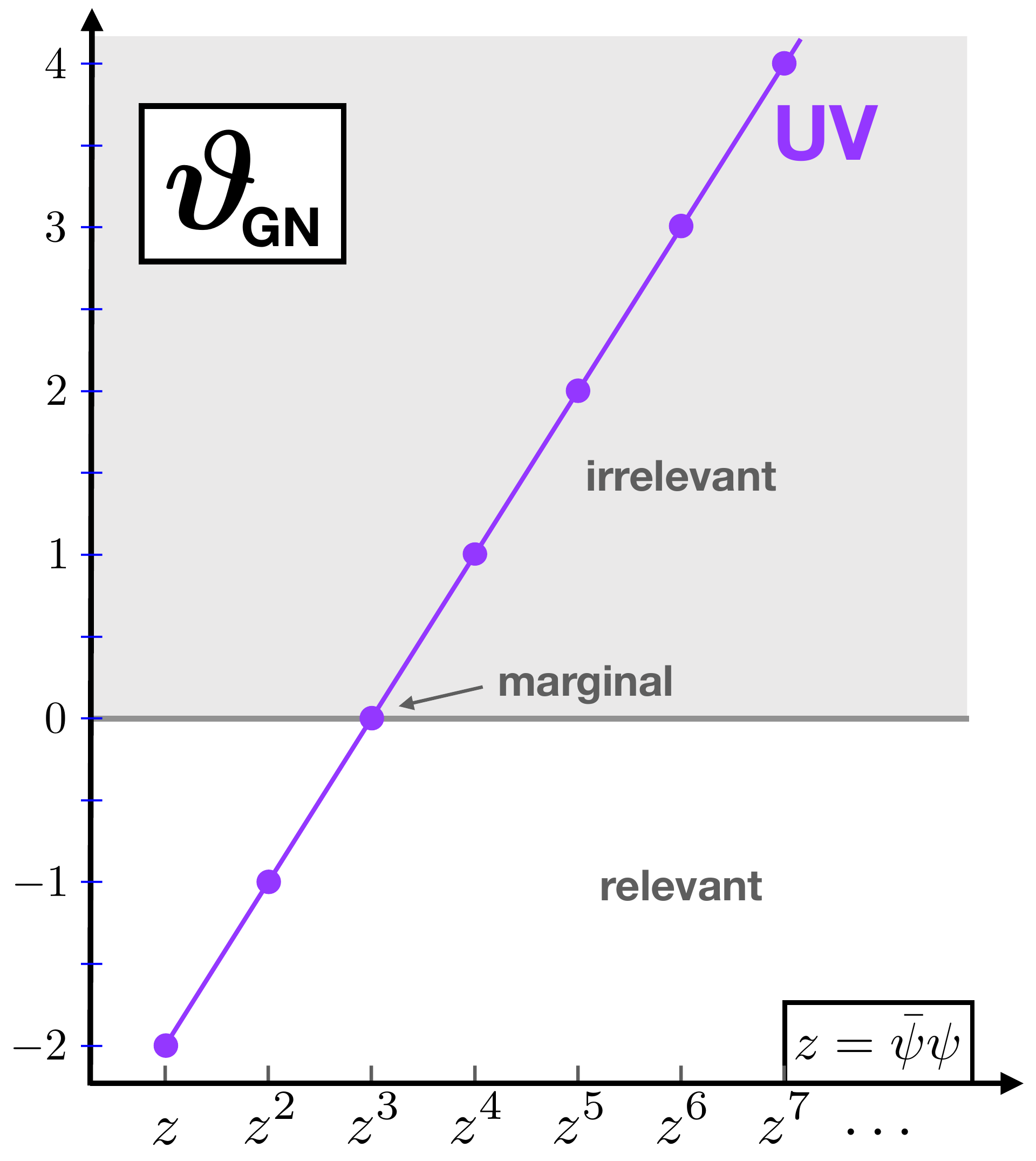}
\caption{Shown are  the universal scaling exponents at the  non-chiral GN fixed points, and their association to   field monomials
(lines are added to guide the eye). Notice that the scaling dimensions agree with those at the interacting GNY fixed point (see Fig.~\ref{fig:pThetas}).}
\label{fig:pThetasGN}
\end{figure}

In Fig.~\ref{fig:pThetasGN}, we show the universal scaling dimensions associated to interaction monomials $\sim z^n$ at the interacting UV fixed point \cite{Cresswell-Hogg:2022lgg},
\be\label{thetaGN}
\vartheta^{\rm GN}_n=-3+n\,.
\ee
Notice that they differ substantially from the scaling dimensions $\vartheta^{\rm GN}_n|_{\rm IR}=-3+2n$ at the free IR fixed point. The infinite set of scaling dimensions of the GN fixed point  (Fig.~\ref{fig:pThetasGN}) agrees with those of scalar field monomials at the GNY  fixed point  (Fig.~\ref{fig:pThetas}, IR line).  The equivalence also persists in the presence of parity-odd interactions. We may associate the scaling dimensions \eqref{thetaGNY}  and the corresponding scaling fields in the GNY theory  to  the scaling dimensions \eqref{thetaGN} and the corresponding scaling operators in the GN theory.

Away from critical points, a common feature is that fermion mass is not generated explicitly by fluctuations, even though chiral symmetry is fundamentally absent. Instead, mass generation remains an exclusively dynamical mechanism, tied to the existence of quantum critical points, thus allowing the generation of mass without the (spontaneous) breaking of a symmetry.

  Further, the role played by the exactly marginal 6F coupling in the GN theory, \eqref{6Fcrit}, is taken over by the exactly marginal cubic scalar coupling in the GNY theory, \eqref{stable}. In either case, the line of fixed points defined by the marginal coupling is finite, and characterised by a critical endpoint where quantum scale symmetry is broken spontaneously.

  Given that the  IR fixed points of the GNY theory \eqref{eq:SGNY} and the  UV fixed points of the GN theory \eqref{eq:SGN} display equivalent conformal scaling dimensions, it strengthens the view that the former can be interpreted as the UV completion of the latter even beyond the chirally symmetric setting. At infinite $N$, the GN theory is characterised by all polynomial interactions $\sim z^n$, while the GNY theory is characterised by all scalar field monomials $\sim \phi^n$, in either case scaling according to \eqref{thetaGN}. However, in addition, the GNY theory also displays the Yukawa interaction, which, at the IR fixed point,  is irrelevant with scaling exponent \eqref{thetaY} (see Fig.~\ref{fig:pThetas}). This scaling exponent is part of the conformal data of the critical GNY theory, yet it appears to have no direct counterpart in the critical GN theory. 
  
This can be understood by noting that, although the Yukawa coupling is the crucial ingredient responsible for generating non-trivial IR scaling behaviour in the GNY model, any explicit dependence on $H$ in the IR limit can actually be removed by rescaling the field $\phi \to H \phi$. From the point of view of bosonisation techniques, it is only the ratio $h^2 / \la{2}$, which is proportional to the 4F coupling, that has physical meaning from the point of view of the GN model \cite{Wilson:1972cf,Eguchi:1976iz,Hasenfratz:1991it,Hands:1991py,ZinnJustin:1991yn,Hands:1992be,Karkkainen:1993ef,Weinberg:1997rv,Rosa:2000ju,Braun:2010tt}.
  
Next, we discuss the equivalence of GN and GNY theories on the level of effective actions. At large-$N$, it has been noted that they are related by a functional Legendre transformation \cite{Weinberg:1997rv}, whereby the quantity $H \phi$  becomes canonically conjugate to the fermion bilinear $\psib_a \psi_a$. Taking the constant-field limit of the  map, we find that the fermionic potential $V$ in \eqref{eq:SGN} is related to the scalar potential $U$ as\footnote{We assume that $U ( \phi )$ is convex.}
\be\label{eq:BFlegendre}
V ( \psib_a \psi_a ) = H \phi \mkern1mu \,\psib_a \psi_a + U ( \phi ),
\ee
where the scalar is considered as a function of the fermion bilinear via $\phi = [ U' ]^{-1} ( -H \psib_a \psi_a )$  \cite{Weinberg:1997rv}. This entails a one-to-one correspondence between the scalar self-couplings $\la{n}$ of the GNY theory and the $2n$-fermion couplings $\la{2n \rm F}$ of the GN theory in which the Yukawa coupling becomes a spectator. 

Expanding \eqref{eq:BFlegendre} around $U'( \phi ) =0$ which corresponds to $\psib_a\psi_a = 0$, and using dimensionless variables, we find concrete bosonisation relations for the fermion mass and all $2n$F fermion couplings in terms of the potential minimum $\sigma_0$ and all scalar self-couplings $\la{n}$ starting with
\begin{align}
\nonumber
\la{2 \rm F} &= \ \ h\, \sigma_0\,, \\ 
\label{eq:4Fcouplings}
\la{4 \rm F} &= -{h^2}/{\la{2}}\,, \\
\nonumber
\la{6 \rm F} &= -{h^3 \la{3}}/{\la{2}^3}\,,\\
\nonumber
\la{8 \rm F} &= \ \ h^4 \left( \la{2} \la{4} - 3 \la{3}^2 \right) / \la{2}^5\,,\\
\nonumber &\ \, \vdots
\end{align}
and similarly for the higher order couplings. A few comments are in order:
\begin{itemize}
\item[(i)] {\it Equivalence along  trajectories and fixed points} \\ 
 Taking the scale derivative of the relations \eqref{eq:4Fcouplings} on either side reproduces the  GN flows   given in \cite{Cresswell-Hogg:2022lgg}  in terms of the GNY flows of this paper, and vice versa.\footnote{It was observed previously that the flow of the four fermion coupling $\la{4 \rm F}$ is identical to the flow of $-h^2 / \la{2}$ in the GNY theory \cite{Braun:2010tt}. Our work establishes that this equivalence is embedded in the   general map  \eqref{eq:BFlegendre}, \eqref{eq:4Fcouplings} and its scale-dependent uplift, with similar relations  for all couplings.} The  findings imply that the Legendre transform  \eqref{eq:BFlegendre} can be lifted to an identity valid at all RG scales  by promoting $V$, $H$ and $U$ to scale-dependent functions $V_k$, $H_k$, $U_k$ under the functional RG flow \eqref{eq:wetterich}.  Notice that the same momentum cutoffs  have been used in the GN and GNY versions of the functional RG flow to achieve the result. This also entails equivalence of universal scaling exponents at quantum critical points, independent of the momentum cutoff.

\item[(ii)] {\it Naturalness of fermion mass}\\
In the absence of explicit fermion mass terms, and irrespective of chiral symmetry, GNY fermions remain massless provided  the  expectation value $\langle H\phi\rangle$ of the scalar field vanishes. Correspondingly, since $H\phi$ is canonically conjugate to the fermion bilinear, the dual GN fermions remain massless as well. If $\langle H\phi\rangle\neq 0$, GNY fermions and their GN counterparts become massive. The flows of the dimensionless mass terms $\partial_t \la{2 \rm F}$ in the GN theory \cite{Cresswell-Hogg:2022lgg} and  $\partial_t(h\, \sigma_0)$  in the GNY theory  \eqref{eq:mF2} are identical after identifiying couplings via the map \eqref{eq:4Fcouplings}. In either theory, fermion mass is found to be technically natural in the sense of `t~Hooft, in that it cannot be switched on by fluctuations directly even if chiral symmetry is absent. 
Hence, the  map on the level of mass terms  now ``explains'' the technical naturalness of fermion mass in GN theories  even if  chiral symmetry is absent \cite{Cresswell-Hogg:2022lgg,Cresswell-Hogg:2022lez} as a consequence of the very existence of a disordered phase  $\langle H\phi\rangle=0$ with  massless fermions in the dual GNY theory, and vice versa. 

\item[(iii)] {\it Critical endpoints and broken hyperscaling}\\
At the endpoints of the IR critical line (Fig.~\ref{fig:phaseDiagram}),  the  exactly marginal  cubic  couplings $\la{3}=\pm\la{3,\rm crit}$  \eqref{crit}  of the GNY theory are mapped by \eqref{eq:4Fcouplings}  onto  their exactly marginal 6F counterparts $\la{6\rm F}=\mp\la{6\rm F, crit}$ \eqref{eq:6Fcrit}, which are the endpoints of the UV critical line of the GN theory. Further, along the entire critical lines of the GN and GNY theories bar their endpoints, the correleation length exponent $\nu$ and the specific heat exponent $\alpha$ are related by the hyperscaling relation \be d\,\nu=2-\alpha\,,\ee  with $d$ the  space-time dimensionality. At the endpoint of the UV line of the GN theory, however,  it has been shown that the hyperscaling relation is violated as a consequence of scale symmetry breaking \cite{Cresswell-Hogg:2022lez}. Given the map \eqref{eq:BFlegendre}, \eqref{eq:4Fcouplings}, we conclude that the hyperscaling relation is equally broken at the endpoints of the IR line of the GNY theory studied in this paper  (Fig.~\ref{fig:phaseDiagram}).

\item[(iv)] {\it Role of the Yukawa coupling}\\
As a final remark, we consider trajectories from the free UV fixed point to the line of interacting IR fixed points (Fig.~\ref{fig:phaseDiagram}). Here, we note that the 4F coupling  in \eqref{eq:4Fcouplings} relates to \eqref{eq:flowX} as $X=\la{4\rm F}/\la{4\rm F}^*$. Since $X\equiv 1$ on UV-IR connecting separatrices \eqref{eq:sep}, it follows that the corresponding 4F coupling does not run and sits at its interacting UV fixed point instead. The reason for this is that  the canonical mass dimension of the Yukawa coupling in the UV is  smoothly rolled-over to  effectively become half the anomalous dimension of the scalar field in the IR, \eqref{eq:flowh}. In particular, this leaves ratios such as $\la{2}(k)/h^2(k)=\la{2,\Lambda}/h^2_{\Lambda}$  and $\la{3}(k)/h^3(k)=\la{3,\Lambda}/h^3_\Lambda$ invariant under the RG flow,  \eqref{eq:sep}, \eqref{la3k}. For the higher order couplings with $n\ge 3$, and provided that $\la{n,\Lambda}/(h_\Lambda)^n=\frac{a_n}{n-3}$, see \eqref{eq:flow-lambda}, \eqref{eq:FPcouplings}, the pattern percolates to the entire effective potential. Hence, even though   the scalar potential $U_k ( \phi )$ and the Yukawa coupling $H_k$ run on UV-IR connecting separatrices, the corresponding fermionic potential $V_k ( \psib_a \psi_a )$ remains strictly at its UV fixed point, because the running is trivialised by the Legendre map. This result also makes the spectator role played by the Yukawa interaction explicit (see Fig.~\ref{fig:pThetas} vs.~Fig.~\ref{fig:pThetasGN}).
\end{itemize}
 
\section{Discussion and conclusions}
\label{DC}

We have investigated  3d GNY theories \eqref{eq:SGNY} in the limit of many fermion flavours with the help of functional renormalisation. These theories are renormalisable in perturbation theory and asymptotically free in the UV. If chiral symmetry is manifest, the theory can develop an interacting fixed point in the IR. Relaxing chiral symmetry, which is the main novelty of this work, the classically relevant cubic scalar self-coupling with canonical mass dimension $\s032$ becomes exactly marginal in the IR due to strong quantum effects \eqref{eq:flow-lambda3}. This new effect has opened up an entire line of interacting and globally well-defined IR fixed points (Fig.~\ref{fig:phaseDiagram}), characterised by the value of the cubic scalar self interaction \eqref{eq:l3freeParam}. 

Moreover, we have  found that the line of fixed points is finite rather than infinite. The reason for this is that the increase of the cubic coupling above a critical  strength triggers the loss of vacuum stability, \eqref{stable}. Exactly at the endpoint of stability, the quantum effective potential becomes flat implying a degenerate  ground state (Fig.~\ref{fig:k0potentialsIR}), and scale symmetry  is broken spontaneously. Most importantly, this leads to the spontaneous generation of a fermion mass  at a quantum critical point while the scalar  remains massless, and  a massless dilaton. It will then be interesting to see if a similar mechanism can generate mass in weakly-coupled particle theories at the end of their conformal window \cite{Litim:2014uca,Bond:2017tbw,Bond:2019npq,Bond:2021tgu,Litim:2023tym}.

Our results are also of interest from the viewpoint of symmetry. It is well known that fermion mass is protected in the presence of  chiral  \eqref{eq:discreteSym} or parity symmetry \eqref{eq:parity}. In our setting, however, chiral symmetry is absent due to scalar cubic interactions switched on at the high scale.  Still, we find that a non-trivial fermion mass cannot be switched on by fluctuations alone, and chiral or parity symmetry are not required to protect fermion mass. Rather, the significantly milder constraint $m_F=0$ at the high scale is already sufficient, e.g.~\eqref{eq:mF2}.  

From the viewpoint of mass generation,  fermion mass only arises when the scalar field develops a non-trivial vacuum expectation value and transitions from a disordered phase into an ordered one. Along the IR line of critical points, and using the scalar mass parameter $\delta\lambda_2$ as a small perturbation away from the fixed point, we find a disordered phase with a vanishing vacuum expectation value and massless fermions for any $\delta\lambda_2>0$. On the other side, we find a massive and ordered phase with  a non-trivial vacuum expectation value provided $\delta\lambda_2<0$. As such, the model provides an example where mass is generated  through a quantum phase transition without breaking any symmetry.

We now discuss how our results for GNY theories \eqref{eq:SGNY} relate to GN theories \eqref{eq:SGN}. At their critical points, theories display equivalent lines of fixed points characterised by an exactly marginal coupling, and equivalent scaling exponents (see Fig.~\ref{fig:pThetas} and~Fig.~\ref{fig:pThetasGN}). Moreover, we have demonstrated that the functional map between the GN and GNY-type actions \cite{Weinberg:1997rv} remains valid along RG flows \eqref{eq:wetterich}, resulting in exact relations between all running couplings, \eqref{eq:4Fcouplings}. This strengthens the view that the critical GNY theories can be interpreted as the UV completion of general GN theories. An intriguing implication of this Legendre map away from critical points is that the naturalness of fermion mass in GN theories \cite{Cresswell-Hogg:2022lgg,Cresswell-Hogg:2022lez} even in the absence of chiral symmetry can now be understood as the consequence of naturalness of fermion mass in their dual GNY theories.

Finally, we   comment on  corrections beyond large  $N$. What  changes  is that the cubic  coupling  $\la{3}$ becomes weakly irrelevant in the critical region, the   line of  fixed points collapses to a single point, and subleading $1/N$ corrections  modify the generation of  mass.  
Still,  the running of the $\la{3}$ remains $1/N$  slow, and
we expect to find a light rather then a massless dilaton as a remnant of broken scale symmetry. 
Further,
 chirally odd interactions  generate a parametrically small fermion mass $\propto 1/N$, and 
 mass generation   in  $\delta\lambda_2$ 
 now takes the form of  a  crossover. 
 With increasing $N$, the crossover becomes increasingly sharp, and practically indistinguishable from a second order quantum phase transition.
 We conclude that key features of theories  at infinite $N$ remain of significance   at finite $N$. A  more detailed study of finite $N$ effects will be reported elsewhere  \cite{Cresswell-Hogg:2024}.

Our results add to the extensive set of critical GNY theories studied in the context of conformal field theory (CFT), see e.g.~\cite{Iliesiu:2015qra,Fei:2016sgs,Iliesiu:2017nrv,Goykhman:2020tsk,Prakash:2022gvb,Erramilli:2022kgp,Herzog:2022jlx,Giombi:2022vnz}. Here, the powerful links between fixed points of the RG and CFTs \cite{Cardy:1996xt} can be exploited to extract conformal data from our results. Further, large-$N$ theories of critical scalars or critical fermions have been shown to be dual to versions of Vassiliev’s higher spin theories under the AdS/CFT conjecture \cite{Klebanov:2002ja,Sezgin:2003pt,Giombi:2012ms}. As such, our study offers new large-$N$ conformal theories involving bosons and fermions simultaneously, including in settings without parity symmetry. It would be intriguing to clarify whether critical GNY theories  have interesting gravity duals, and how the spontaneous breaking of scale symmetry becomes visible in conformal correlators \cite{Maldacena:2011jn,Maldacena:2012sf}.

\acknowledgments
This work is  supported by the Science and Technology Facilities Council (STFC)  under the Consolidated Grant T/X000796/1, and  was performed in part during the 2023 Aspen Center for Physics  workshop {\it Emergent Phenomena of Strongly-Interacting Conformal Field Theories and Beyond}, which is supported by National Science Foundation grant PHY-2210452.

\bibliography{GNY}
\bibliographystyle{JHEP}

\end{document}